\documentclass[aps,prb,twocolumn,showpacs,groupedaddress]{revtex4-1}
\usepackage{amsmath}
\usepackage{amssymb}
\usepackage{graphicx}
\usepackage{dcolumn}
\usepackage{bm}
\usepackage{subfigure}

\newcommand{\ph}{\textit{p}-H$_2$}

\begin{document}

\title{Superfluidity of metastable bulk glass \textit{para}-hydrogen at low
      temperature}
\author{O. N. Osychenko$^{\rm a}$, R. Rota$^{\rm b}$ and J. Boronat$^{\rm a}$}
\affiliation{
$^{\rm a}$ Departament de F\'{i}sica i Enginyeria Nuclear, Universitat Polit\`{e}cnica 
de Catalunya, Campus Nord B4-B5, E-08034, Barcelona, Spain \\
$^{\rm b}$  Dipartimento di Fisica, Universit\`a di Trento and INO-CNR BEC Center, 
I-38123 Povo, Trento, Italy
} 

\begin{abstract}

Molecular \textit{para}-hydrogen (\ph) has been proposed theoretically as a
possible candidate for superfluidity, but the eventual superfluid
transition is hindered by its crystallization. In this work, we study a
metastable non crystalline phase of bulk \ph \ by means of the Path
Integral Monte Carlo method in order to investigate at which temperature
this system can support superfluidity. By choosing accurately the initial
configuration and using a non commensurate simulation box, we have been
able to frustrate the formation of the crystal in the simulated system and
to calculate the temperature dependence of the one-body density matrix and
of the superfluid fraction. We observe a transition to a superfluid phase
at temperatures around 1 K. The limit of zero temperature is also studied
using the diffusion Monte Carlo method. Results for the energy, condensate
fraction, and structure of the  metastable liquid phase  at $T=0$ are
reported and compared with the ones obtained for the stable solid phase.

%\textbf{Keywords:} solid helium; first-principles; shear modulus; high pressure  

\end{abstract}

\maketitle

\section{Introduction}

Superfluidity and Bose-Einstein condensation (BEC) have been stunningly shown in
metastable dilute alkali gases, magnetically confined at ultralow
temperatures.~\cite{sandro} The extreme diluteness of these gases allows for the
achievement of BEC with an almost full occupation of the zero-momentum
state that has been possible to observe and measure quite easily. This
contrasts with the difficulties encountered in the measure of the
condensate fraction in liquid $^4$He, which amounts to only 8\% at the
equilibrium density.~\cite{glyden0} However, liquid $^4$He is a stable superfluid below the
lambda transition $T_\lambda=2.17$ K and is therefore a more easily
accessible system. Before the blowup produced in the field of quantum fluids by the first 
experimental realization of BEC gases, liquid helium was the only paradigm
of a superfluid. From long time ago, there has been great interest in the
search of superfluid condensed phases other than liquid helium.
Spin-polarized atomic deuterium and tritium are predicted to be fermionic 
and bosonic liquids, respectively, in the limit of zero
temperature.~\cite{panoff,ivana}
However, their experimental study has proven to be very elusive due to their
high recombination rate, and only the case of atomic hydrogen, whose 
ground state is a gas, has been experimentally driven to its BEC
state.~\cite{klepner}    
The next candidate for superfluidity is molecular hydrogen, which has been studied for a long
time.~\cite{silvera} This seems a priori an optimal system due to its very light mass but
it crystallizes at relatively high temperature as a
consequence of the intensity of its intermolecular attraction, without
exhibiting any superfluid transition in the liquid phase. In the
present work, we study the properties of metastable liquid or glass
molecular hydrogen at very low temperatures using quantum Monte Carlo
methods.

In 1972, Ginzburg and Sobyanin~\cite{Ginzburg72} proposed that any Bose liquid
should be superfluid below a certain temperature $T_{\lambda}$,
unless it solidifies at temperature $T_f$ higher than
$T_{\lambda}$. To give a first estimation of $T_{\lambda}$, 
they used the ideal Bose gas theory, obtaining
\begin{equation}
T_{\lambda} = 3.31 \, \frac{\hbar^2}{g^{2/3} m k_B} \rho^{2/3} \ ,
\label{eq:Tlambdaideal}
\end{equation}
where $m$ is the atomic mass, $g$ is the spin degeneracy, $k_B$ is
the Boltzmann constant and $\rho$ is the density of the system.
Ginzburg and Sobyanin proposed molecular \textit{para}-hydrogen (\ph) as a plausible 
candidate for superfluidity: being a
spinless boson ($g=1$) with a small mass, \ph \  should
undergo a superfluid transition at a relatively high
temperature (according to Eq. (\ref{eq:Tlambdaideal}), $T_{\lambda}
\simeq 6$ K).

The estimation of $T_{\lambda}$, given by Eq. (\ref{eq:Tlambdaideal}), 
is clearly inaccurate in the case of dense liquids because it cannot 
account for the
observed dependence of $T_{\lambda}$ with the density. In fact, 
$T_{\lambda}$ slightly decreases in liquid $^4$He when $\rho$ increases, a
manifestly opposite behavior to the increase with $\rho^{2/3}$ given by the
ideal gas formula (\ref{eq:Tlambdaideal}). In order to
provide a more reasonable estimation of $T_{\lambda}$, Apenko~\cite{Apenko99}
proposed a phenomenological prescription for the superfluid
transition, similar to the Lindemann criterion for classical crystal melting.
In this way, he was able to take into
account quantum decoherence effects due to the strong
interatomic potential and to relate the critical temperature for
superfluidity with the mean kinetic energy per particle above the
transition. For \ph, he concluded that $T_{\lambda}$
should vary between $1.1$ K and $2.1$ K, depending on the density of
the system.

Superfluid \ph \ is not observed in a stable form because it
crystallizes at temperature $T_f = 13.8$ K, which is significantly higher
than the expected $T_{\lambda}$.
Several studies about crystal nucleation in \ph \  have been performed 
in order to understand if the liquid can
enter a supercooled phase, i.e., a metastable phase in which the
liquid is cooled below its freezing temperature without forming a
crystal. Maris \textit{et al.}~\cite{Maris83}
calculated the rate $\Gamma(T)$ of homogeneous nucleation of the
solid phase from the liquid as a function of the temperature $T$,
showing a maximum of $\Gamma$ around $T = 7$ K and a rapid
decrease at lower temperature. This suggests that, if it would be
possible to supercool the liquid through the range where $\Gamma$
is large, one might be able to reach a
low-temperature region where the liquid is essentially stable.
However, recent experiments have indicated that, even at $T \sim
9$ K, the rate of crystal growth is so high that the liquid phase
freezes quickly into a metastable polymorph crystal.~\cite{kuhnel11}

Even though several supercooling techniques have been proposed 
to create a metastable liquid phase in bulk 
\ph,~\cite{Maris87,Vorobev00,Grisenti06} none of them has proven so
far to be
successful and no direct evidence of superfluidity has been
detected. However, there are evidences of superfluidity in several  
spectroscopic studies of  small doped \ph \ clusters.  
In 2000, Grebenev \textit{et al.}~\cite{Grebenev00} analyzed the rotational
spectra of a linear carbonyl sulfide (OCS) molecule surrounded by 14 to 16
\ph \  molecules absorbed in a larger helium  droplet, which
fixes the temperature of the cluster. 
When \ph \ is immersed in a $^4$He droplet ($T= 0.38$ K), 
the measured spectra shows a peak indicating the
excitation of angular momentum around the OCS axis. On the other hand, 
if the small
\ph \  cluster is put inside a colder $^4$He-$^3$He droplet
($T = 0.15$ K), the peak disappears: the OCS
molecule is then able to rotate freely inside the hydrogen cluster,
pointing to the superfluidity of the surrounding \ph \ molecules. 
These results have been confirmed in a later experiment on small
\ph \  clusters doped with carbon dioxide.~\cite{Li10} From a
precise analysis of the rotational spectra, it has been possible
to measure the effective momentum of inertia of these small
systems, and thus of their superfluid fraction $\rho_s$, providing
a clear evidence of superfluidity in clusters made up of $N \le
18$ \ph \  molecules. These clusters are too small to extract reliable
predictions of a metastable liquid phase and larger clusters would be
desirable. To this end, Kuyanov-Prozument and Vilesov~\cite{Prozument08}  have been
able to stabilize liquid clusters with an average size of
$N \approx 10^4$ \ph \  molecules down to temperature $T= 2$ K, 
but they do not see any evidence of superfluidity. Other attempts of
producing liquid \ph \ well below $T_f$  ($T=1.3$ K) are based on the generation
of continuous hydrogen filaments of macroscopic
dimensions.~\cite{Grisenti06}

The search for a superfluid \ph \ phase has been intense also from the
theoretical point of view. The rather simple radial form of the \ph-\ph
interaction and the microscopic accuracy achieved by quantum Monte Carlo
methods have stimulated a long-standing effort for devising possible scenarios
where supercooled \ph \ could be studied. In practically all the cases, the
search is focused on systems of reduced dimensionality or in finite
systems. Path integral Monte Carlo (PIMC) simulations of \ph \ films adsorbed on a surface with
impurities observed superfluidity for some arrangements of these
impurities,~\cite{Gordillo97} but these results were posteriorly questioned by
other PIMC studies.~\cite{Boninsegni05H2} In a one-dimensional channel, like
the one provided experimentally by narrow carbon nanotubes, it has been
predicted a stable liquid phase in the limit of zero temperature.~\cite{h21d}
The largest number of theoretical works have been devoted to the study of small
clusters, both pure~\cite{Sindzingre91,Mezzacapo06,Mezzacapo07,
Khairallah07,Mezzacapo08,ester,guardiola,cuervo06,cuervo08,cuervo09} and doped with impurities.~\cite{Kwon02,
Paesani05,Kwon05} All these simulations show that \ph \  
becomes superfluid below a certain temperature $T=1$-$2$ K and that the 
superfluid fraction depends on the number of molecules
of the cluster. When the cluster becomes larger than a certain molecular number 
($N > 18$-$25$), solid-like structures are observed and the superfluidity 
vanishes.   

In the present work, we deliver a PIMC study of a metastable glass/liquid
phase at very low temperature. Our main purpose has been to determine for
the first time at
which temperature this metastable phase becomes superfluid and the value of
the superfluid density and condensate fraction  close to this temperature. The simulations are
carried out following schedules which are similar to the ones used in a
recent study of a glass $^4$He  phase evolving from a normal to a
superfluid state (superglass).~\cite{glass4he} Our results show that this
transition temperature is $T\simeq 1$ K, a value that is close to the Apenko
estimation~\cite{Apenko99} and also close to the values observed in
simulations of small clusters. As a  complementary aspect, we address the
calculation of the equation
of state of the metastable liquid \ph \ phase in the limit of zero
temperature using the diffusion Monte Carlo (DMC) method. The simulation of
the liquid phase  in this limit is easier than at finite temperature and 
therefore DMC is able to provide accurate information on its main energetic
and structure properties.    

The rest of the paper is organized as follows. In Sec. II, we introduce
the quantum Monte Carlo methods used in the study, the DMC and PIMC
methods,
and report specific details on how the simulations are carried out. Sec.
III contains the results of the equation of state, structure properties,
and condensate fraction of metastable liquid \ph \ at zero temperature.
PIMC results at finite temperature are reported in Sec. IV, and finally
the main conclusions of the present work are discussed in Sec. V.

\section{Quantum Monte Carlo methods}

The H$_2$ molecule, which  is composed of two hydrogen
atoms linked by a covalent bond, is spherically symmetric in the
\textit{para}-hydrogen state (total angular momentum zero).  
The energy scale involved in electronic excitations   
($\sim 10^{5}$ K) is orders of magnitude larger than the intermolecular
one  ($\sim 10^{1}$ K), thus modelling the \ph-\ph \ interaction by means of a 
radial  pair-potential and 
considering the molecules as point-like turns out to be justified upon the
condition  of low or moderate pressures.  In this work, we have chosen the
well-known and commonly used semiempirical Silvera-Goldman pair 
potential.~\cite{SilveraGoldman} 
This potential has proved to be accurate at low  temperature and
in the pressure regimes in which we are interested.

The study in the limit of zero temperature has been  performed
with  the DMC method. DMC is a first-principles method which can access 
exactly the ground state of bosonic systems.
It is a form of Green's Function Monte Carlo
which samples the projection of the ground state from the initial configuration
with the operator $\exp{\left[-(\mathcal{H}-E_0)\tau\right]}$. Here, $\mathcal{H}$
is the Hamiltonian
\begin{equation}
\mathcal{H} = -\frac{\hbar^2}{2m} \sum_{i=1}^{N} {\bm \nabla}_i^2 +
\sum_{1=i<j}^{N} V(r_{ij}) \ ,
\label{hamiltonian}
\end{equation}
$E_0$ is a norm-preserving adjustable constant and $\tau$
is the imaginary time. The simulation is 
performed by advancing in $\tau$ via a combination of diffusion, drift and branching
steps on walkers $\bm{R}$ (sets of $3N$ coordinates) representing the wavefunction of 
the system.~\cite{dmcboro}  The imaginary-time evolution of the walkers is
``guided'' during the drift stage by a guiding wavefunction $\phi_G$, which
is usually a good guess for the  wavefunction of the system. This function
contains basic ingredients of the system as its symmetry, phase and
expected behaviors at short and long distances according to its
Hamiltonian. Technically,  $\phi_G$ allows  importance sampling and thus
reduces the variance of the ground-state estimations. It is straightforward 
to show that for the Hamiltonian  $\mathcal{H}$ and any operator commuting with 
it, the expectation value  is computed exactly within statistical error.
Other diagonal operators which do not fulfill this condition require of a special
treatment, known as pure estimation,~\cite{pures} which leads also for this case to
unbiased results.

The phase of the system is imposed within the typical imaginary-time length
by the guiding wave function. This property of the DMC method is here a key
point if we are pursuing a investigation of the properties of 
the metastable liquid \ph \ phase. Then, for the liquid phase $\phi_G$ is
taken in a Jastrow form
\begin{equation}
\phi_G (\bm{R}) = \prod_{1=i<j}^{N} f(r_{ij}) \ ,
\label{jastrow}
\end{equation}
with a two-body correlation function~\cite{f2reatto}
\begin{equation}
f(r) = \exp \left[ - \frac{1}{2} \, \left( \frac{b}{r} \right)^5 
-\frac{L}{2} \, \exp \left[ - \left( \frac{r - \lambda}{\Lambda} \right)^2
\right] \right] \ .
\label{f2}
\end{equation}
In order to compare the results obtained for the liquid phase with the ones
corresponding to the stable hcp solid we have carried out some simulations
with a guiding wave function of Nosanow-Jastrow type
\begin{equation}
\phi_G^s (\bm{R}) = \prod_{1=i<j}^{N} f(r_{ij}) \, \prod_{i=1}^{N}
g(r_{iI}) \ ,
\label{nosanow}
\end{equation}
the set $\{ \bm{r}_I \}$ being the lattice points of a perfect hcp lattice.
Optimal values for the parameters entering Eq. (\ref{f2}) are  $b=3.68$ \AA,
$L=0.2$, $\lambda=5.24$ \AA, and $\Lambda=0.89$ \AA \ for the liquid phase,
and $b=3.45$ \AA, $L=0.2$, $\lambda=5.49$ \AA, and $\Lambda=2.81$ \AA \ for
the solid one. The Nosanow term is chosen in Gaussian form, $g(r)=\exp
(-\gamma r^2)$.  The density dependence of 
the parameters in the Jastrow term is small, and 
neglected in practice when used in DMC, whereas  the Nosanow term 
parameter $\gamma$ is optimized for the whole range of densities. 
We have used 256 and 180 particles per
simulation box for the liquid and hcp solid phases, respectively. The number of
walkers and time-step have been adjusted to reduce any bias coming from
them to the level of the statistical noise.

At finite temperature $T$, the microscopic description of the quantum system is
made in terms of the thermal density matrix,
$\rho_N(\bm{R}',\bm{R};\beta)=\langle \bm{R}' \vert
e^{-\beta \mathcal{H}} \vert \bm{R} \rangle$, with $\beta=(k_B T)^{-1}$.
The partition function $Z$, which allows for a full description of the
properties of a given system, satisfies the relation
\begin{equation}
Z = \rm{Tr}(e^{-\beta \mathcal{H} }) \simeq \int \prod_{i=1}^{M} d \bm{R}_i \, 
\rho_N \left({\bf R}_i,{\bf R}_{i+1};\varepsilon\right) \ ,
\label{PartitionFunction}
\end{equation} 
that relies on the convolution property of the density matrix. In Eq.
(\ref{PartitionFunction}), $\varepsilon=\beta/M$ and the boundary condition 
$R_{M+1} = R_1$ applies. The remarkable feature of Eq.
(\ref{PartitionFunction}), on which PIMC is based, is that one can access to
information at a temperature $T$ by convoluting density matrices at higher
temperature $M T$.~\cite{rmp_ceperley}

PIMC describes the quantum $N$-body system
considering $M$ different configurations $\bm{R}_j$ of the same system,
whose sequence constitutes a path in imaginary time. 
This means that the $N$-body quantum system is mapped onto a
classical system of $N$ ring polymers, each one composed by $M$ beads. The
different beads can be thought as a way to describe the delocalization of
the quantum particle due to its zero-point motion.
For sufficiently large $M$, one recovers the
high-temperature  density matrix, where it is legitimate to separate the kinetic
contribution from the potential one (primitive action). In this way, 
it is possible by applying Eq. (\ref{PartitionFunction}) to reduce
the systematic error due to the analytical approximation for $\rho_N$ below
the statistical uncertainty. However, 
the primitive action is too simple to study extreme quantum matter and a
better choice for the action is fundamental to reduce both the complexity of
the calculation and ergodicity issues. To this end, we have used a high-order
Chin action~\cite{chin,Sakkos09}  to obtain an accurate  estimation of the
relevant physical quantities with reasonable numeric effort even in the low
temperature regime, where the simulation becomes harder  due to the large
zero-point motion of particles. We have analyzed the dependence of the \ph
\  energy on the parameter $\varepsilon$ and determined an optimal value
$\varepsilon=1/60$ K$^{-1}$ for which the bias coming from the use of a
finite  $\varepsilon$ value is smaller than the characteristic statistical
noise.

A relevant issue one has to deal with when approaching the low
temperature limit with PIMC simulations arises from the indistinguishable
nature of the particles. In the path integral formalism, the exchanges
between $L$ different particles are represented by long ring polymers
composed by $L \times M$ beads. If we study a bosonic system like \ph, the
indistinguishability of the particles does not affect the positivity of the
integrand function in Eq. (\ref{PartitionFunction}) and the symmetry of
$Z$ can be recovered by the  sampling of permutations between the
ring polymers. In the present study, we have used the  
Worm Algorithm~\cite{BoninsegniWorm} which provides a very efficient
sampling in permutation space.

The key aspect of the worm algorithm is to work in an extended configuration 
space, containing not only the usual diagonal configurations made up of ring 
polymers, but also off-diagonal configurations which are characterized  by the
presence of an open polymer (defined as the {\it worm}).  By working with
off-diagonal configurations, it is possible to sample the  bosonic permutations
by means of single-particle movements, like the {\it swap} update,  whose
acceptance rate can be made comparable to that of the other  updates in the
sampling of polymers. In order to fulfill this condition, it is important  to
optimize two parameters of the worm algorithm. The first of them is $C$, which
regulates  the acceptance probability of the movements switching from diagonal
to  off-diagonal configuration and vice versa. In our simulations, we choose $C$
to get the number of sampled off-diagonal  configurations to about 65-70\% of
total number of configurations.  In this way, the system is allowed to spend
enough time both in off-diagonal  configurations, where the sampling of
permutations is done, and in diagonal  configurations, where relevant
observables such as the energy or the  superfluid density are evaluated. The
second one is $M_s$, which  is the number of beads rebuilt in the swap update.
The parameter $M_s$ has to be chosen as a  compromise between a small value,
which would make difficult the search  of the partner of the worm in the swap
movement, and a large value which  would make difficult the reconstruction of
the polymer once the partner has been chosen.  In our simulations, we use $M_s$
which maximizes the  acceptance rate of the swap update: the typical value of
$M_s$  is about 10\% of the number of beads $M$.

\section{Zero-temperature results}

We have calculated the main properties of the metastable liquid and stable
hcp solid phases of \ph. Our main goal has been to know the properties of
a hypothetical bulk liquid phase and compare them with the ones of the
stable solid. In order to achieve reliable estimations of liquid \ph \ it
is crucial to work with a guiding wave function of liquid type, as we have
discussed in the preceding Section. Within the typical imaginary-time
length of our simulations we have not seen the formation of any crystal
structure, i.e.,  no signatures of
Bragg peaks in the structure function $S(k)$ have been registered so far.

\begin{figure}
\centerline{
\includegraphics[width=0.9\linewidth]{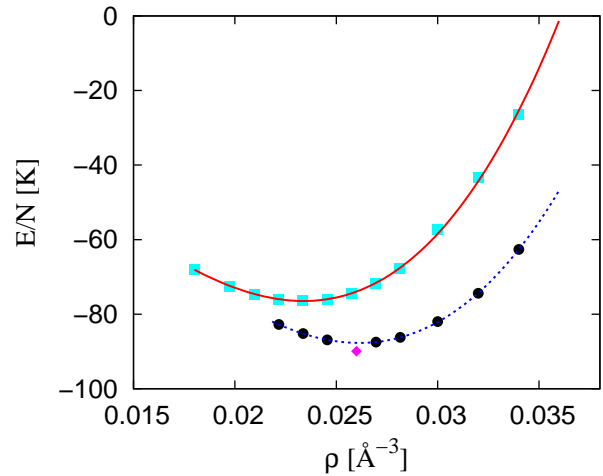}}%
\caption{DMC energies per particle as a function of the density. Squares
and circles correspond to the liquid and solid phases, respectively. Solid
and dashed lines are the polynomial fits to the DMC energies for the liquid
and solid, respectively. The diamond is the experimental energy of hcp molecular
hydrogen from Ref. \onlinecite{phexperimental}}  
\label{fig:ener0t}
\end{figure}

\begin{figure}[b]
\centerline{
\includegraphics[width=0.9\linewidth]{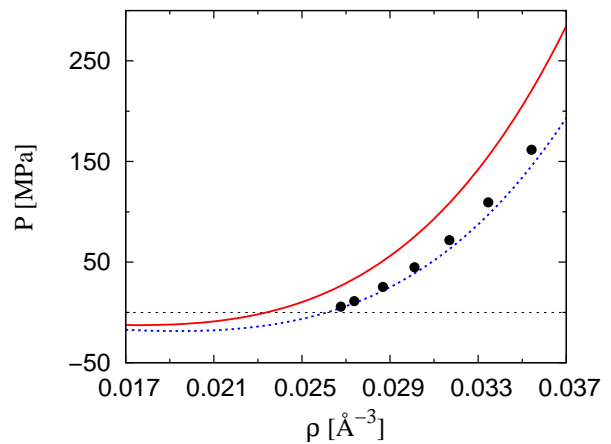}}
\caption{Pressure of the liquid (solid line) and solid (dashed line) \ph \ phases as a 
function of the density. Experimental points for the solid phase~\cite{phexperimental2} 
are show as solid circles. 
}  
\label{fig:pres0t}
\end{figure}

In Fig. \ref{fig:ener0t}, we plot the DMC energies per particle of metastable
liquid \ph \ as a function of the density. For comparison, we also report
the results obtained for the hcp crystal phase. Our hcp energies are in
close agreement with the ones reported in Ref. \onlinecite{pederiva} using
the same Silvera-Goldman potential. In the figure, we also show the 
experimental estimation at $T=0$ K from Ref. \onlinecite{phexperimental},
$E/N=-89.9$ K, that lies a bit below of our results. This is again in
agreement with previous DMC results~\cite{pederiva} which show that the experimental
energy is, in absolute value, underestimated and overestimated by the
Silvera-Goldman and Buck potential,~\cite{buck} respectively. Our results for both
phases are well reproduced by the polynomial law
\begin{equation}
\frac{E}{N} = \left( \frac{E}{N} \right)_0 + A \left( \frac{\rho-\rho_0}{\rho_0}
\right)^2 + B \left( \frac{\rho-\rho_0}{\rho_0} \right)^3 \ ,
\label{eqstate}
\end{equation}  
$(E/N)_0$ and $\rho_0$ being the equilibrium energy per particle and
equilibrium density, respectively. These equations of state are shown in
Fig. \ref{fig:ener0t} with lines. The optimal parameters of the fits are:
$\rho_0=0.026137(20)$ \AA$^{-3}$, $(E/N)_0=-87.702(37)$ K, $A=235(2)$ K,
$B=140(10)$ K for the solid, and $\rho_0=0.023386(40)$ \AA$^{-3}$, 
$(E/N)_0=-76.465(51)$ K, $A=188(1)$ K, $B=131(10)$ K for the liquid. As
expected, our DMC results shows that the solid phase is the stable one with
a difference in energy per particle at the respective equilibrium points of
$\sim 10$ K, the equilibrium density of the liquid being $\sim 10$ \%
smaller than the solid one. The same trend was observed in a DMC simulation
of two-dimensional \ph, but there the differences were significantly
smaller.~\cite{claudi2d} It is worth noticing that about one half of the 
energy difference in the bulk systems
comes from the decrease of the kinetic energy per particle going from the
liquid to the solid: at density $\rho=0.03$ \AA$^{-3}$, it amounts to $93.3(1)$
and to $89.5(1)$ K for the liquid and solid, respectively.

\begin{figure}[t]
\centerline{
\includegraphics[width=0.9\linewidth]{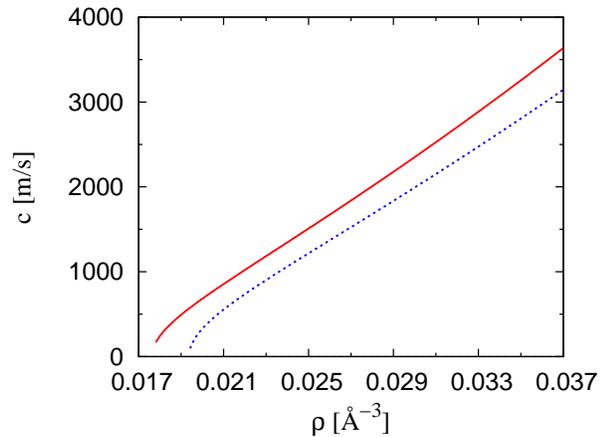}}%
\caption{Speed of sound of the liquid (solid line) and solid (dashed line) \ph \ phases as a 
function of the density.  
}  
\label{fig:velocitat0}
\end{figure}

From the equations of state (\ref{eqstate}), it is easy to deduce the
pressure of the system at any density using the relation $P(\rho)= \rho^2 [
d (E/N) / d \rho]$. The results obtained for metastable liquid and stable
solid phases are shown in Fig. \ref{fig:pres0t}. As one can see, at a given
density the pressure of the liquid is larger than the one of the solid
mainly because of the different location of the equilibrium densities ($P=0$).
The results for the solid are compared with experimental data from Ref.
\onlinecite{phexperimental2} showing a good agreement especially for not
very large pressures. The density at which the function $P(\rho)$ has a
zero slope defines the spinodal point; beyond this limit the system is no
more thermodynamically stable as a homogeneous phase. At this point, the
speed of sound $c(\rho)= [ m^{-1} (d P / d \rho)]^{1/2}$ becomes zero.
Results for $c(\rho)$ are shown for both phases in Fig.
\ref{fig:velocitat0}. The speed of sound decreases when the density is
reduced and drops to zero at the spinodal point: ($\rho_c=0.0176(1)$
\AA$^{-3}$,
$P_c=-12.6(5)$ MPa) and   ($\rho_c=0.0193(1)$ \AA$^{-3}$, $P_c=-18.5(5)$ MPa) for
liquid and solid, respectively.

\begin{figure}[]
\begin{center}
\includegraphics[width=0.73\linewidth]{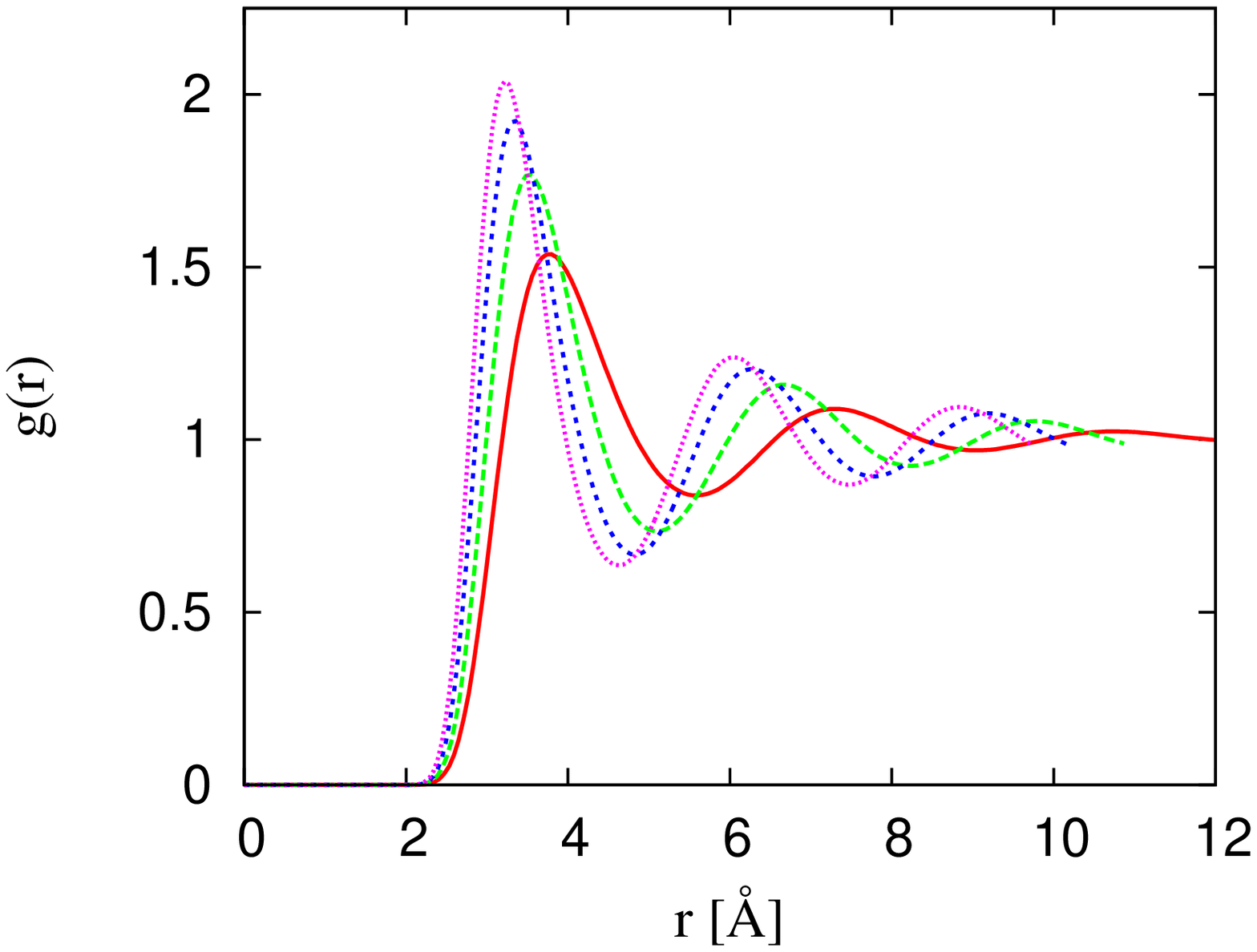} 
\\
\includegraphics[width=0.7\linewidth]{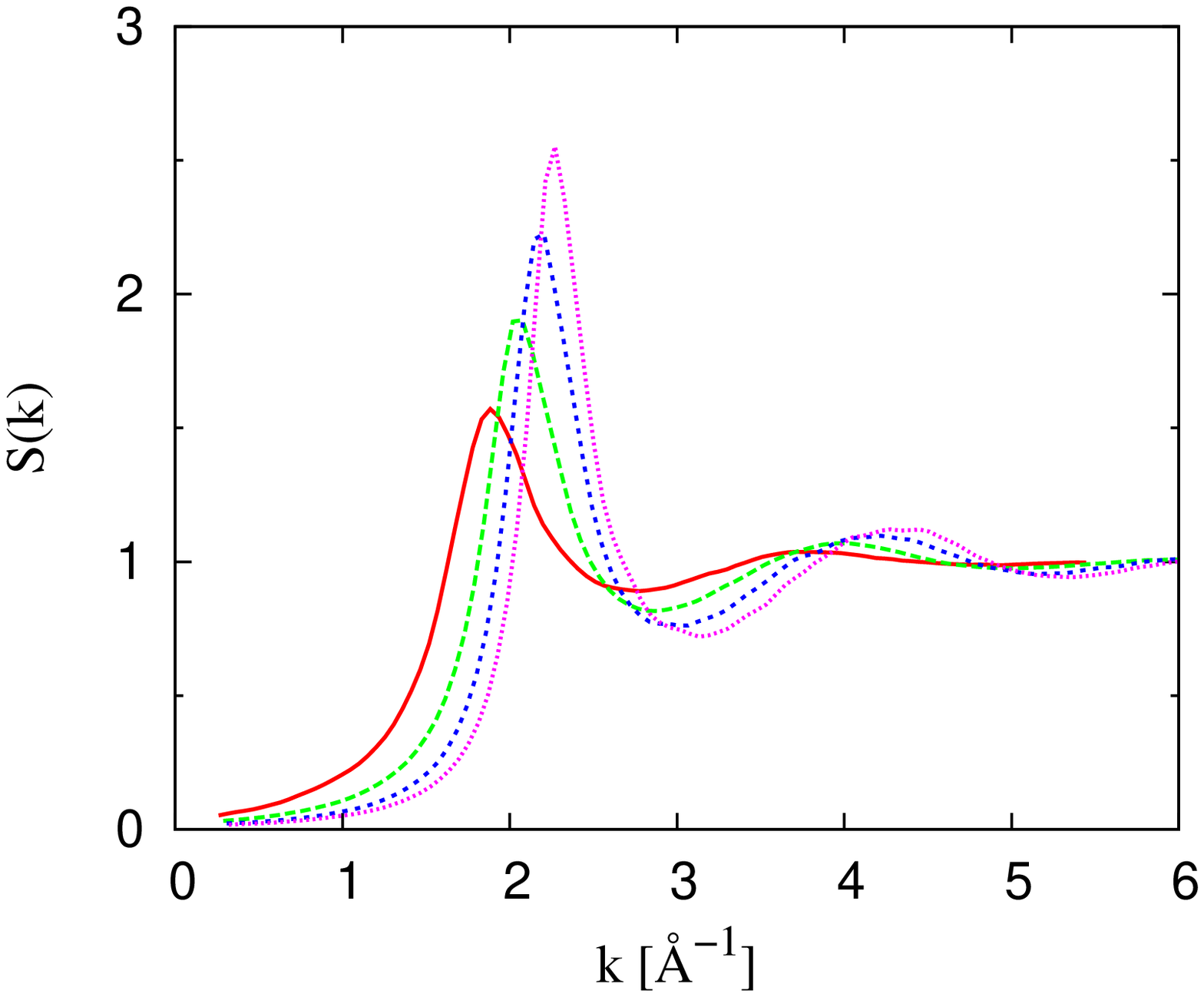}
\end{center}
\caption{\textit{Top panel:} Two-body radial distribution function of the
liquid \ph \ phase at different densities: solid, long-dashed, 
short-dashed, and dotted lines stand for densities 
$\rho=0.0180$, $0.0245$, $0.0300$, and $0.0340$ \AA$^{-3}$, respectively.
\textit{Bottom panel:} Static structure factor of the liquid phase. Same
densities and notation than in the top panel.  
}  
\label{fig:grskdens}
\end{figure}

\begin{figure}[b]
\centerline{
\includegraphics[width=0.9\linewidth]{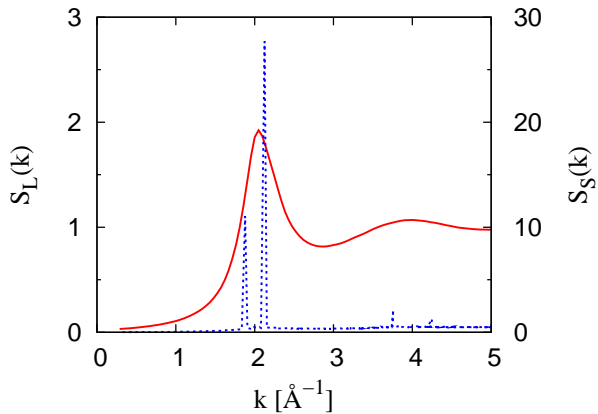}}%
\caption{Static structure function of liquid and solid \ph \ at  
density $\rho=0.0245$ \AA$^{-3}$. The result for the liquid $S_L(k)$ (left
scale) is shown with a solid line; the one for the hcp solid $S_S(k)$
(right scale) with a dashed line.
}  
\label{fig:skcomp}
\end{figure}

DMC produces also accurate results for the structure of the bulk system. In
Fig. \ref{fig:grskdens}, we show results for the two-body radial distribution
function $g(r)$ of the liquid \ph \ phase for a set of densities. This
function is proportional to the probability of finding two molecules
separated by a distance $r$. Increasing the density, the main peak becomes
higher and moves to shorter interparticle distances; at least three peaks
are observed. All these features point to the picture of a very dense
liquid, with much more structure than in stable liquid $^4$He. In the same
Fig. \ref{fig:grskdens}, we show results for the static structure factor
$S(k)$, related to $g(r)$ by a Fourier transform. As one can see, the main
peak increases quite fast with the density suggesting a highly structured
metastable liquid. Nevertheless, we have not observed within the scale of
the simulations the emergence of any Bragg peak which would point to
formation of crystallites in the simulation box. In Fig. \ref{fig:skcomp},
we illustrate the comparison between $S(k)$ for the liquid and solid
systems at a density $\rho=0.0245$ \AA$^{-3}$, close to the equilibrium
density of the liquid. The difference is the one expected between a liquid
and a solid: oscillating function towards one at large $k$ for the liquid
and a sequence of Bragg peaks, corresponding to the hcp lattice, for the
solid.

\begin{figure}[t]
\centerline{
\includegraphics[width=0.9\linewidth]{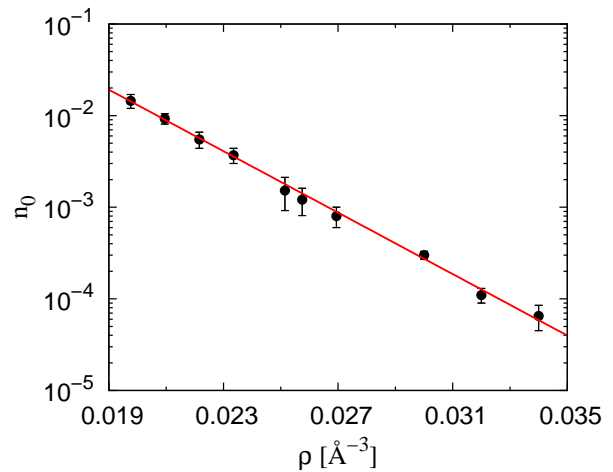}}%
\caption{Condensate fraction of metastable liquid \ph \ as a function of
the density. The points are the DMC results and the line is an exponential
fit to them.}  
\label{fig:condensate}
\end{figure}

One of the most relevant properties of a superfluid is the mean occupation
of the zero-momentum state, i.e., the condensate fraction $n_0$. As it is
well known, $n_0$ can be obtained from the asymptotic behavior of the
one-body density matrix $\rho_1(r)$,
\begin{equation}
n_0= \lim_{r \rightarrow \infty}  \rho_1(r)   \ , 
\label{condensate}
\end{equation}
with $\rho_1(r)$ being obtained as the expectation value of the operator
\begin{equation}
\left \langle \frac{ \Phi({\bf r}_1,\ldots,{\bf r}_i+{\bf r},\ldots,{\bf
r}_N )}{\Phi( {\bf r}_1,\ldots,{\bf r}_N ) }  \right \rangle \ .
\label{obdm2}
\end{equation}
DMC results for the condensate fraction of liquid \ph  \ as a function of
the density,
obtained using the extrapolated estimator (there are no reliable pure estimators
for non-diagonal operators), are shown in Fig. \ref{fig:condensate}. The
decrease of $n_0$ with the density is well described by an exponential
decay (line in the figure). The strong interactions induced by the deep
attractive potential well produce a big depletion of the condensate state.
At the equilibrium density, our estimation for the condensate fraction is
$n_0=0.0037(7)$. This value is more than one order of magnitude smaller
than the measured condensate fraction~\cite{glyden0} of liquid $^4$He at equilibrium
($0.08$).

\section{Superfluid transition temperature}

One of the main goals of our work has been to determine the temperature at
which a disordered  phase of \ph \ becomes eventually
superfluid. Recently, a similar approach has been used to study the
superfluid properties of a glass phase of $^4$He,~\cite{glass4he} a system 
that has been named
superglass and that it has been argued to be related with some of the
effects observed in torsional oscillator experiments on solid $^4$He.

The first difficulty we have to deal with when investigating
computationally a disordered phase of \ph \  at
low temperature is to provide a good equilibration of the system.
The PIMC method, indeed, is aimed at studying the thermodynamic
properties of quantum systems at thermal equilibrium. On the contrary, 
our purpose here is to study a configuration different from
the one of minimum free energy, which for \ph \  at low
temperature is the crystalline one.
In order to do that, it is fundamental to choose thoroughly 
the dimensions of the simulation box and the number of
particles, which must not be commensurate with any crystalline
lattice. Also, it is important to choose a good initial
configuration which evolves, as the Monte Carlo simulation goes
on, towards a non crystalline phase which remains metastable for a
number of Monte Carlo steps large enough to get a good statistics of
the relevant quantities of the system. In this equilibration
process, special attention must be paid to the thermalization
of the polymers used within the PIMC formalism. 
A bad choice of initial conditions  may cause the 
evolution of the system towards a configuration where the
polymers representing each molecule are not allowed to spread and thus 
are not able
to describe properly the zero-point motion of the molecules. This
eventuality may represent a serious problem in our simulation,
since we are mainly interested in the investigation of the
superfluid properties of \ph.

To check whether an equilibration scheme is efficient or not, it
is important to monitor how the numerical estimations of the
physical quantities change with the number of Monte Carlo steps.
If we see that, as the simulation goes on, the computed variables
do not show any evident trend but fluctuate around a certain
value, we can conclude that the system has reached the
metastability. To check if this eventual metastable phase is
crystalline or not, we can calculate the static structure factor
$S(k)$ and observe if it presents the Bragg peaks typical of a
crystal configuration.

We have used different technical schemes to get the desired metastable
glass/liquid configuration. In many cases, we were not able to stabilize
this phase and after some time the liquid froze. Finally, we managed to
devise a successful approach that is based in the following two steps. 
In the first part, we perform the simulation of a fictitious 
system of quantum particles with a mass equal to the one of the \ph \  molecule,
but interacting through the $^4$He-$^4$He Aziz potential.~\cite{Aziz}
Compared with the H$_2$ Silvera-Goldman pair interaction, the Aziz potential does
not present a so deep attractive well and thus it is not able to 
freeze the system. Once this fictitious system is equilibrated, we change the Aziz
interaction  by the Silvera-Goldman one and equilibrate again towards the
metastable \ph \ phase. 
At all the temperatures we consider in our study, we have verified that 
the superfluid properties of \ph \ do not depend on the fact that 
permutations are allowed or not in the first step of the equilibration.
To test this equilibration scheme, we have performed a simulation
of $N = 100$ \ph \  molecules interacting through the
Silvera-Goldman potential,~\cite{SilveraGoldman} inside a cubic box at the
equilibrium density of the liquid phase at zero temperature (see Sec. III),
$\rho = 0.0234 \, \textrm{\AA}^{-3}$. 
For a preliminary test, we choose to perform the PIMC
simulation at temperature $T = 10$ K, that is an intermediate
temperature below the freezing temperature, where the liquid
phase should be unstable, but above the estimated superfluid
transition temperature, in order to make the simulation easier. 
It is worth noticing that after equilibration the glass phase is better
sampled, and thus crystallization is avoided, when the center of mass of 
all the polymers are moved simultaneously and accepted or not collectively
by a single Metropolis step.

\begin{figure}
 \begin{center}
  \includegraphics[width=0.9\linewidth,angle=0]{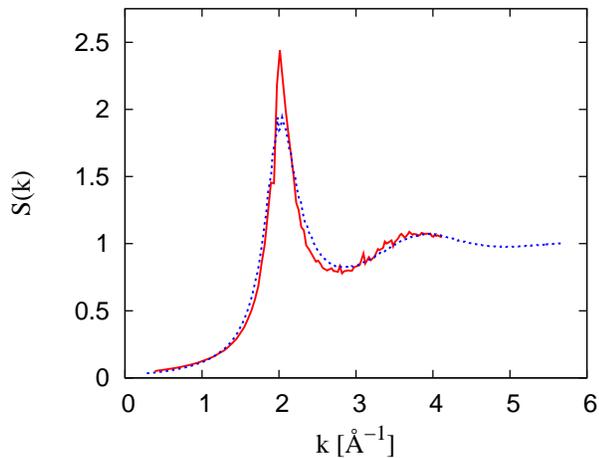}
 \end{center}
\caption{Static structure factor $S(k)$ of metastable 
bulk \ph \  at $T = 10$ K obtained with PIMC (solid
line). This result is compared with the static structure factor of
liquid \ph \  at zero temperature obtained with DMC (dotted
line)} 
\label{fig:skT10}
\end{figure}

In Fig. \ref{fig:skT10}, we have plotted the static structure factor $S(k)$
of  the metastable phase and compared it with the same quantity computed
for the  liquid phase with DMC. We can see that the curve obtained
at $T = 10$ K  presents the first peak at the same $k$ as the $S(k)$ of the
liquid and follows the same behavior up to the second maximum
which is at  $k \simeq 4 \, \textrm{\AA}^{-1}$. Even though the PIMC
calculation gives a peak which is higher and narrower than the peak
obtained with DMC, and indicates that the PIMC configurations are slightly
more structured than the DMC ones, we can  conclude that our
equilibration scheme is able to create a metastable liquid phase, at least
in the range of intermediate temperature below the freezing point $T_f$ and
above the expected superfluid transition $T_{\lambda}$.

\begin{figure}
 \begin{center}
  \includegraphics[width=0.9\linewidth,angle=0]{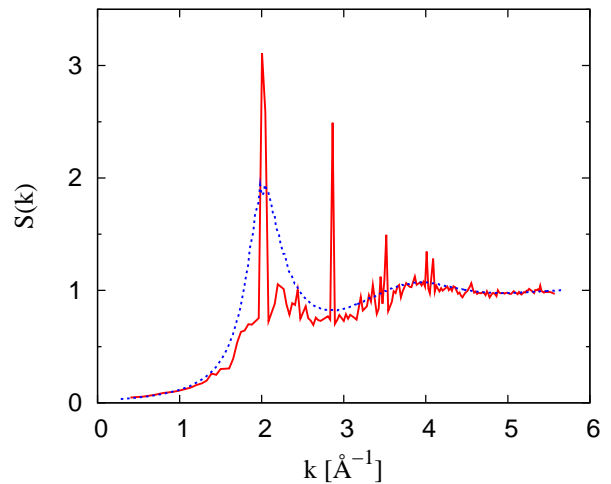}
   \end{center}
\caption{Static structure factor $S(k)$ of metastable 
bulk \ph \  at $T = 2$ K obtained with PIMC (solid
line). For comparison, we also show  the $S(k)$
of liquid \ph \  at zero temperature
obtained with DMC (dotted line). } 
\label{fig:skT2}
\end{figure}

Since our main purpose is to localize the superfluid
transition of this non crystalline phase, it is worth to test this equilibration 
scheme at lower temperatures, closer to the expected $T_{\lambda}$. For
this reason, we have performed a simulation with $N = 90$  \ph \  
molecules at the same density, $\rho = 0.0234 \,
\textrm{\AA}^{-3}$,  but at a lower temperature, $T = 2$ K.
Once the mean value of the energy was stable, we computed the static structure
factor $S(k)$. The result is shown in Fig.
\ref{fig:skT2}, in comparison with the static structure factor of the
zero-temperature liquid. As we can see, $S(k)$ obtained in the PIMC simulation
presents narrow maxima in the range of small $k$ and is 
different from the typical $S(k)$ of a liquid phase. However,
these maxima tend to disappear at higher $k$ and their height is
much lower than the height of the Bragg peaks appearing in the
$S(k)$ of a crystal. This indicates that the system simulated with
PIMC has relaxed to a glass phase, which is structured at
short range but lacks of the long-range coordination typical of
the crystal structures.

\begin{figure}
 \begin{center}
  \includegraphics[width=0.9\linewidth,angle=0]{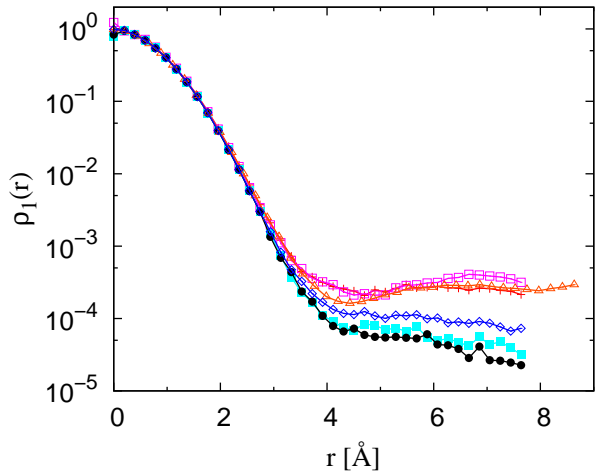}
 \end{center}
\caption{One-body density matrix $\rho_1(r)$ for  glass
\ph \ at density $\rho = 0.0234 \, \textrm{\AA}^{-3}$ and
at different temperatures. 
$T = 0.7 \, {\rm K}$ (crosses), $T = 1.0 \, {\rm K}$  (open squares), 
$T = 1.0 \, {\rm K}$ and $N=130$ (triangles),
$T = 1.2 \, {\rm K}$ (diamonds),
$T = 1.5 \, {\rm K}$ (full squares),
 and $T = 2.0 \, {\rm K}$ (circles). 
%Statistical
%errors, if not displayed, are smaller than the symbol
%size.
}
\label{fig:obdmh2}
\end{figure}

Even if a glassy configuration can
make the diffusion of particles harder, the lack of long range
coordination makes possible the appearance of off-diagonal long
range order and it is worth to study the superfluid properties of
this phase. To do that, we have studied the temperature dependence of the
one-body density matrix $\rho_1(r)$. 
The first simulation has been performed at 
$T = 2$ K, using $N = 90$ \ph \  molecules
inside a cubic box at density $\rho = 0.0234 \, \textrm{\AA}^{-3}$. 
The result
for $\rho_1(r)$ obtained in this simulation is shown in Fig. \ref{fig:obdmh2}:
at $T = 2$ K, we can clearly see an exponential decay of
$\rho_1(r)$ at large $r$, indicating that Bose-Einstein
condensation is not present in the system. 
In fact, we have noticed
that in this calculation the swap update (the update responsible for bosonic exchanges in our
PIMC sampling)  has a very low acceptance rate 
and does not
allow the formation of long-permutation cycles with a non-zero
winding number. Nevertheless, one can think that the low acceptance
of the swap update is a consequence of the difficulties in the
sampling of the coordinates due to the strength of the
intermolecular potential. We may therefore suspect that the system
remains stuck in a configuration without permutation because of
sampling issues. To be sure about our result, we have performed
another simulation of the same system but starting from an initial
configuration presenting a non-zero winding number. To create this
initial configuration, we have allowed particles to permute even
in the fictitious simulation used to equilibrate the system. When
we start the PIMC simulation of \ph \  from the permutated
configuration, we see that the percentage of particles involved in
bosonic exchanges tends to decrease and, at the end of the
equilibration, the system has relaxed to a phase presenting zero
winding number, i.e., the superfluid density is zero. This last result 
confirms our conclusion that the
\ph \  glass we simulate is not superfluid at $T = 2$ K.

In Fig. \ref{fig:obdmh2}, we have also shown $\rho_1(r)$ estimated at other
temperatures. At each of the temperatures studied, we
have performed simulations starting both from a permutated and a
non permutated configuration, observing that once  the system
has been equilibrated the results for $\rho_1(r)$ do not depend
on the initial configuration. From the comparison of the curves at
different temperatures, we can easily see a change of the behavior
of $\rho_1$ at large $r$ as the temperature decreases: this
indicates that, at temperatures close to $T = 1$ K, the system
presents a transition to a superfluid phase presenting
off-diagonal long range order. The condensate fraction at low
temperature is $n_0 \sim 3 \times 10^{-4}$, appreciably smaller than the
DMC estimation for the liquid phase at $T=0$. The observation of a finite
value for the condensate fraction at $T=1$ K agrees with the measure of
a finite value for the superfluid density. Our results show that 
the superfluid density, derived from the winding number estimation, is zero
within our numerical uncertainty for $T>1.2$ K and at $T=1$ K is already
$\rho_s/\rho = 0.36 \pm 0.08$. 
In order to give a more precise estimation of the superfluid
transition temperature $T_{\lambda}$, it would be necessary to
perform a pertinent finite-size scaling study. However, 
the achievement of the metastable state is quite hard for 
systems made up of more than $\sim 100$ molecules.
In our study, we have been able to stabilize the amorphous phase for a
system made up of 130 \ph \ molecules at $T=1$ K. The result for 
$\rho_1(r)$ is also shown in Fig. \ref{fig:obdmh2}. We notice that this
result is in agreement with the one calculated for the smaller system of 90
\ph \ molecules at the same temperature, supporting our conclusion that the
system has undergone a superfluid transition. At higher temperatures,
instead, the system made up of 130 \ph \ molecules relaxes  to a
crystalline phase and it has not been possible to calculate $\rho_1(r)$
for the amorphous configuration. This result seems to indicate that the
disordered phase is somehow ``more stable" at lower temperature, when the
\ph \ molecules begin to permutate. A similar behavior has also been found in
PIMC simulations of small \ph \  clusters.~\cite{Mezzacapo06}

\section{Conclusions}

We have carried out extensive quantum Monte Carlo calculations of \ph \ at
temperatures well below its solidification point. Our interest has been to
know better the properties of the metastable liquid/glass phase at very
low temperatures and to determine where the superfluid transition is
expected to appear. In the limit of zero temperature we have used the DMC
method, which is a very efficient tool to sample this metastable
phase through the use of a proper guiding wave function. The results point to a very
structured liquid with a large depletion of the condensate fraction,
significantly larger than in stable liquid $^4$He.

Our estimation of $T_\lambda \sim 1 \, \textrm{K}$ is slightly smaller than the 
prediction obtained using a  
phenomenological approach~\cite{Apenko99} for which, at the density
$\rho = 0.0234 \, \textrm{\AA}^{-3}$ studied in our simulations, the
 transition temperature is estimated to be $T_{\lambda} \sim 1.7 \,
\textrm{K}$. It is also interesting to notice that our result 
for $T_{\lambda}$ is
quite close to the temperatures at which, according to PIMC simulations,
superfluid effects should appear in small \ph \  
clusters.~\cite{Sindzingre91,Mezzacapo06} These calculations show that
clusters made of $N \le 20$ \ph \  molecules exhibit a non-zero
superfluid fraction below $T \sim 2$ K. This transition temperature depends
on the dimension of the cluster, decreasing when the number of molecules
increases. However, it is difficult to make hypothesis on the superfluid
behavior of large enough \ph \  systems from the simulation of small
clusters, because the calculated superfluid fraction $\rho_s$ is
significantly depressed when the number of molecules becomes $N \ge 30$.
This unexpected behavior of $\rho_s$ with $N$ has been explained by relating
the changes in the superfluid properties to structural changes that make
the molecules arrange according to a solidlike configuration when the
dimension of the cluster becomes large. In our simulation of bulk glass \ph, 
we have been able to frustrate crystallization  with an efficient equilibration
of the system and to measure
finite values of both the condensate fraction and superfluid density.

\begin{acknowledgments}
The authors acknowledge partial financial support from the  
DGI (Spain) Grant No.~FIS2011-25275 and Generalitat de Catalunya 
Grant No.~2009SGR-1003.
\end{acknowledgments}


\begin{thebibliography}{30}

\bibitem{sandro} L. Pitaevskii and S. Stringari, \textit{Bose-Einstein
Condensation} (Clarendon Press, Oxford, 2003).

\bibitem{glyden0}  H.R. Glyde, S.O. Diallo, R.T. Azuah, O. Kirichek, and J.W. Taylor, 
Phys. Rev. B \textbf{83}, 100507 (2011).

\bibitem{panoff} R. M. Panoff and J. W. Clark, Phys. Rev. B \textbf{36},
5527 (1987). 

\bibitem{ivana}  I. Be\v{s}li{\'c}, L. Vranje\v{s} Marki{\'c}, and J.
Boronat,  Phys. Rev.  B \textbf{80}, 134506 (2009).  

\bibitem{klepner} D. G. Fried, T. C. Killian, L. Willmann, D. Landhuis, S. C.
Moss, D. Kleppner, and T. J. Greytak, Phys. Rev. Lett. \textbf{81}, 3811
(1998).  

\bibitem{silvera} I. F. Silvera, Rev. Mod. Phys. \textbf{52}, 393 (1980).
 
\bibitem{Ginzburg72} V. L. Ginzburg and A. A. Sobyanin, JEPT Lett.
\textbf{15}, 242 (1972).

\bibitem{Apenko99} S. M. Apenko, Phys. Rev. B \textbf{60}, 3052 (1999).

\bibitem{Maris83} H. J. Maris, G. M. Seidel, and T. E. Huber, 
J. Low Temp. Phys. \textbf{51}, 471 (1983).

\bibitem{kuhnel11} M. K\"uhnel, J. M. Fern\'andez,  G. Tejeda, A. Kalinin, 
S. Montero, and R. E. Grisenti, Phys. Rev. Lett. \textbf{106}, 245301 (2011).

\bibitem{Maris87} H. J. Maris, G. M. Seidel, and F. I. B. Williams, 
Phys. Rev. B \textbf{36}, 6799 (1987).

\bibitem{Vorobev00} V. S. Vorob'ev and S. P. Malyshenko, 
J. Phys.: Condens. Matter \textbf{12}, 5071 (2000).

\bibitem{Grisenti06} R. E. Grisenti, R. A. Costa Fraga, N. Petridis, R.
D{\"o}rner, and J. Deppe, Europhys. Lett. \textbf{73}, 540 (2006).

\bibitem{Grebenev00} S. Grebenev, B. Sartakov, J. P. Toennies, and A. F. Vilesov, 
Science \textbf{289}, 1532 (2000).

\bibitem{Li10} H. Li, R. J. Le Roy, P. N. Roy, and A. R. W. McKellar, 
Phys. Rev. Lett. \textbf{105}, 133401 (2010).

\bibitem{Prozument08} K. Kuyanov-Prozument and A. F. Vilesov, Phys. Rev. Lett. 
\textbf{101}, 205301 (2008).

\bibitem{Gordillo97} M. C. Gordillo and D. M. Ceperley, Phys. Rev. Lett. 
\textbf{79}, 3010 (1997).

\bibitem{Boninsegni05H2} M. Boninsegni, New J. Phys. \textbf{7}, 78 (2005).

\bibitem{h21d} M. C. Gordillo, J. Boronat, and J. Casulleras,
Phys. Rev. Lett. \textbf{85}, 2348 (2000).

\bibitem{Sindzingre91} P. Sindzingre, D. M. Ceperley, and M. L. Klein, 
Phys. Rev. Lett. \textbf{67}, 1871 (1991).
 
\bibitem{Mezzacapo06} F. Mezzacapo and M. Boninsegni, 
Phys. Rev. Lett. \textbf{97}, 045301 (2006).

\bibitem{Mezzacapo07} F. Mezzacapo and M. Boninsegni, Phys. Rev. A 
\textbf{75}, 033201 (2007).
 
\bibitem{Khairallah07} S. A. Khairallah, M. B. Sevryuk, D. M. Ceperley, and J. P. Toennies, 
Phys. Rev. Lett. \textbf{98}, 183401 (2007).

\bibitem{Mezzacapo08} F. Mezzacapo and M. Boninsegni, Phys. Rev. Lett. 
\textbf{100}, 145301 (2008).

\bibitem{ester} E. Sola and J. Boronat, J. Phys. Chem. A \textbf{115}, 7071
(2011).
 
\bibitem{guardiola}  R. Guardiola and  J. Navarro, Cent. Eur. J. Phys.
\textbf{6}, 33 (2008).

\bibitem{cuervo06} J. E. Cuervo and P. N. Roy, J. Chem. Phys. 
\textbf{125}, 124314 (2006).

\bibitem{cuervo08} J. E. Cuervo and P. N. Roy, J. Chem. Phys. 
\textbf{128}, 224509 (2008).

\bibitem{cuervo09} J. E. Cuervo and P. N. Roy, J. Chem. Phys. 
\textbf{131}, 114302 (2009).


\bibitem{Kwon02} Y. Kwon and K. B. Whaley, Phys. Rev. Lett. \textbf{89}, 273401 (2002).
  
\bibitem{Paesani05} F. Paesani,  R. E. Zillich,  Y. Kwon, and K. B. Whaley, 
J. Chem. Phys. \textbf{122}, 181106 (2005). 

\bibitem{Kwon05} Y. Kwon and K. B. Whaley, J. Low Temp. Phys. \textbf{140}, 227 (2005).

\bibitem{glass4he} M. Boninsegni, N. Prokof'ev, and B, Svistunov,
Phys. Rev. Lett. \textbf{96}, 105301 (2006).
 
\bibitem{SilveraGoldman} I. F. Silvera and V. V. Goldman, J. Chem. Phys.
\textbf{69}, 4209 (1978).

\bibitem{dmcboro}  J. Boronat and J. Casulleras, Phys. Rev. B \textbf{49},
8920 (1994).
 
\bibitem{pures} J. Casulleras and J. Boronat, Phys. Rev. B \textbf{52},
3654 (1995).

\bibitem{f2reatto} L. Reatto, Nucl. Phys. \textbf{A328}, 253 (1979).

\bibitem{rmp_ceperley} D. M. Ceperley, Rev. Mod. Phys. \textbf{67}, 279
(1995).


\bibitem{chin} S. A. Chin and C. R. Chen, J. Chem. Phys. \textbf{117}, 1409
(2002).

\bibitem{Sakkos09} K. Sakkos, J. Casulleras, and J. Boronat, J. Chem. Phys.
\textbf{130}, 204109 (2009).

\bibitem{BoninsegniWorm} M. Boninsegni, N. V. Prokof'ev, and B. V. Svistunov, 
Phys. Rev. E \textbf{74}, 036701 (2006).

\bibitem{pederiva}  F. Operetto and F. Pederiva, Phys. Rev. B \textbf{73},
184124 (2006).

\bibitem{phexperimental}  O. Schnepp, Phys. Rev. A \textbf{2}, 2574 (1970).

\bibitem{buck} U. Buck, F. Huisken, A. Kohlhase, D. Otten, and J. Schaefer, J.
Chem. Phys. \textbf{78}, 4439 (1983);  
M. J. Norman, R. O. Watts, and U. Buck, J. Chem. Phys. \textbf{81},
3500 (1984).

\bibitem{claudi2d} C. Cazorla and J. Boronat, Phys. Rev. B \textbf{78},
134509 (2008). 

\bibitem{phexperimental2}  A. Driessen, J. A. de Waal, and I. F. Silvera, 
J. Low Temp. Phys. \textbf{34}, 255 (1979).

\bibitem{Aziz} R. A. Aziz, F. R. W. McCourt, and C. C. K. Wong, 
Mol. Phys. \textbf{61}, 1487 (1987).



\end{thebibliography}
\end{document}